\font\tenfrakturb=eufb10
\font\tenfraktur=eufm10
\font\tenmsbm=msbm10
\font\sevenfrakturb=eufb7
\font\sevenfraktur=eufm7
\font\sevenmsbm=msbm7
\font\fivefrakturb=eufb5
\font\fivefraktur=eufm5
\font\fivemsbm=msbm5
\newfam\bgothicfam
\newfam\gothicfam
\newfam\msbmfam
\textfont\bgothicfam = \tenfrakturb \scriptfont\bgothicfam=\sevenfrakturb
\scriptscriptfont\bgothicfam=\fivefrakturb
\textfont\gothicfam = \tenfraktur \scriptfont\gothicfam=\sevenfraktur
\scriptscriptfont\gothicfam=\fivefraktur
\textfont\msbmfam = \tenmsbm \scriptfont\msbmfam=\sevenmsbm
\scriptscriptfont\msbmfam=\fivemsbm

\def\Bbb{\tenmsbm\fam\msbmfam}

\catcode`@=11
\def\renewcounter#1{\@definecounter{#1}\@ifnextchar[{\@newctr{#1}}{}}

\documentstyle[twoside]{article}
\catcode`\@=11
\long\def\@makefntext#1{
\protect\noindent \hbox to 3.2pt {\hskip-.9pt  
$^{{\eightrm\@thefnmark}}$\hfil}#1\hfill} 

\def\@makefnmark{\hbox to 0pt{$^{\@thefnmark}$\hss}} 
	
\def\ps@myheadings{\let\@mkboth\@gobbletwo
\def\@oddhead{\hbox{}
\rightmark\hfil\eightrm\thepage}   
\def\@oddfoot{}\def\@evenhead{\eightrm\thepage\hfil
\leftmark\hbox{}}\def\@evenfoot{}
\def\sectionmark##1{}\def\subsectionmark##1{}}
\oddsidemargin=\evensidemargin
\newcounter{sectionc}\newcounter{subsectionc}\newcounter{subsubsectionc}
\renewcommand{\section}[1] {\vspace{12pt}\addtocounter{sectionc}{1} 
\setcounter{subsectionc}{0}\setcounter{subsubsectionc}{0}\noindent 
	{\tenbf\thesectionc. #1}\par\vspace{5pt}}
\renewcommand{\subsection}[1] {\vspace{12pt}\addtocounter{subsectionc}{1} 
	\setcounter{subsubsectionc}{0}\noindent 
	{\bf\thesectionc.\thesubsectionc. {\kern1pt \bfit #1}}\par\vspace{5pt}}
\renewcommand{\subsubsection}[1] {\vspace{12pt}\addtocounter{subsubsectionc}{1}
	\noindent{\tenrm\thesectionc.\thesubsectionc.\thesubsubsectionc.
	{\kern1pt \tenit #1}}\par\vspace{5pt}}
\newcommand{\nonumsection}[1] {\vspace{12pt}\noindent{\tenbf #1}
	\par\vspace{5pt}}
\newcounter{appendixc}
\newcounter{subappendixc}[appendixc]
\newcounter{subsubappendixc}[subappendixc]
\renewcommand{\thesubappendixc}{\Alph{appendixc}.\arabic{subappendixc}}
\renewcommand{\thesubsubappendixc}
	{\Alph{appendixc}.\arabic{subappendixc}.\arabic{subsubappendixc}}
\renewcommand{\appendix}[1] {\vspace{12pt}
        \refstepcounter{appendixc}
        \setcounter{figure}{0}
        \setcounter{table}{0}
        \setcounter{lemma}{0}
        \setcounter{theorem}{0}
        \setcounter{corollary}{0}
        \setcounter{definition}{0}
        \setcounter{equation}{0}
        \renewcommand{\thefigure}{\Alph{appendixc}.\arabic{figure}}
        \renewcommand{\thetable}{\Alph{appendixc}.\arabic{table}}
        \renewcommand{\theappendixc}{\Alph{appendixc}}
        \renewcommand{\thelemma}{\Alph{appendixc}.\arabic{lemma}}
        \renewcommand{\thetheorem}{\Alph{appendixc}.\arabic{theorem}}
        \renewcommand{\thedefinition}{\Alph{appendixc}.\arabic{definition}}
        \renewcommand{\thecorollary}{\Alph{appendixc}.\arabic{corollary}}
        \renewcommand{\theequation}{\Alph{appendixc}.\arabic{equation}}
        \noindent{\tenbf Appendix \theappendixc #1}\par\vspace{5pt}}
\newcommand{\subappendix}[1] {\vspace{12pt}
        \refstepcounter{subappendixc}
        \noindent{\bf Appendix \thesubappendixc. {\kern1pt \bfit #1}}
	\par\vspace{5pt}}
\newcommand{\subsubappendix}[1] {\vspace{12pt}
        \refstepcounter{subsubappendixc}
        \noindent{\rm Appendix \thesubsubappendixc. {\kern1pt \tenit #1}}
	\par\vspace{5pt}}
\newcommand{\textlineskip}{\baselineskip=13pt}
\newcommand{\smalllineskip}{\baselineskip=10pt}
\def\eightcirc{
\begin{picture}(0,0)
\put(4.4,1.8){\circle{6.5}}
\end{picture}}
\def\eightcopyright{\eightcirc\kern2.7pt\hbox{\eightrm c}} 
\newcommand{\copyrightheading}[1]
	{\vspace*{-2.5cm}\smalllineskip{\flushleft
	{\footnotesize Modern Physics Letters A, #1}\\
	{\footnotesize $\eightcopyright$\, World Scientific Publishing
	 Company}\\
         }}
\newcommand{\pub}[1]{{\begin{center}\footnotesize\smalllineskip 
	Received #1\\
	\end{center}
        }}

\def\abstracts#1#2#3{{
        \centering{\begin{minipage}{4.5in}\baselineskip=10pt\footnotesize
        \parindent=0pt #1\par 
        \parindent=15pt #2\par
        \parindent=15pt #3
        \end{minipage}}\par}} 

\newcommand{\bibit}{\nineit}
\newcommand{\bibbf}{\ninebf}
\renewenvironment{thebibliography}[1]
         {\frenchspacing
         \ninerm\baselineskip=11pt
         \begin{list}{\arabic{enumi}.}
         {\usecounter{enumi}\setlength{\parsep}{0pt}     
         \setlength{\leftmargin 12.7pt}{\rightmargin 0pt} 
         \setlength{\itemsep}{0pt} \settowidth
         {\labelwidth}{#1.}\sloppy}}{\end{list}}
\newcounter{itemlistc}
\newcounter{romanlistc}
\newcounter{alphlistc}
\newcounter{arabiclistc}

\newcommand{\fcaption}[1]{
         \refstepcounter{figure}
         \setbox\@tempboxa = \hbox{\footnotesize Fig.~\thefigure. #1}
         \ifdim \wd\@tempboxa > 5in
           {\begin{center}
         \parbox{5in}{\footnotesize\smalllineskip Fig.~\thefigure. #1}
            \end{center}}
        \else
             {\begin{center}
             {\footnotesize Fig.~\thefigure. #1}
              \end{center}}
        \fi}
\newcommand{\tcaption}[1]{
        \refstepcounter{table}
        \setbox\@tempboxa = \hbox{\footnotesize Table~\thetable. #1}
        \ifdim \wd\@tempboxa > 5in
           {\begin{center}
        \parbox{5in}{\footnotesize\smalllineskip Table~\thetable. #1}
            \end{center}}
        \else
             {\begin{center}
             {\footnotesize Table~\thetable. #1}
              \end{center}}
        \fi}
\def\@citex[#1]#2{\if@filesw\immediate\write\@auxout
        {\string\citation{#2}}\fi
\def\@citea{}\@cite{\@for\@citeb:=#2\do
        {\@citea\def\@citea{,}\@ifundefined
        {b@\@citeb}{{\bf ?}\@warning
        {Citation `\@citeb' on page \thepage \space undefined}}
        {\csname b@\@citeb\endcsname}}}{#1}}
\newif\if@cghi
\def\cite{\@cghitrue\@ifnextchar [{\@tempswatrue
        \@citex}{\@tempswafalse\@citex[]}}
\def\citelow{\@cghifalse\@ifnextchar [{\@tempswatrue
        \@citex}{\@tempswafalse\@citex[]}}
\def\@cite#1#2{{$\null^{#1}$\if@tempswa\typeout
        {IJCGA warning: optional citation argument 
        ignored: `#2'} \fi}}

\def\pmb#1{\setbox0=\hbox{#1}
        \kern-.025em\copy0\kern-\wd0
        \kern.05em\copy0\kern-\wd0
        \kern-.025em\raise.0433em\box0}

\def\fnt#1#2{\footnotetext{\kern-.3em
        {$^{\mbox{\scriptsize #1}}$}{#2}}}
\def\fpage#1{\begingroup
\voffset=.3in
\thispagestyle{empty}\begin{table}[b]\centerline{\footnotesize #1}
       \end{table}\endgroup}
\def\runninghead#1#2{\pagestyle{myheadings}
\markboth{{\protect\footnotesize\it{\quad #1}}\hfill}
{\hfill{\protect\footnotesize\it{#2\quad}}}}
\font\tenrm=cmr10
\font\tenit=cmti10 
\font\tenbf=cmbx10
\font\bfit=cmbxti10 at 10pt
\font\ninerm=cmr9
\font\nineit=cmti9
\font\ninebf=cmbx9
\font\eightrm=cmr8

\def\qed{\hbox{${\vcenter{\vbox{  
   \hrule height 0.4pt\hbox{\vrule width 0.4pt height 6pt
   \kern5pt\vrule width 0.4pt}\hrule height 0.4pt}}}$}}

\begin{document}
\def\bh{${\Bbb R}^2\times {\Bbb S}^2\>$}
\runninghead{Yu. P. Goncharov \& N. E. Firsova}
{Increase of the Hawking radiation for spinor particles}
\normalsize\textlineskip
\thispagestyle{empty}
\setcounter{page}{2399}
\copyrightheading{Vol. 16, No. 37 (2001) 2399--2407}
\vspace*{0.88truein}
\fpage{2399}
\centerline{\bf INCREASE OF THE HAWKING RADIATION FOR}
\vspace*{0.035truein}
\centerline{\bf SPINOR PARTICLES FROM SCHWARZSCHILD BLACK HOLES}
\vspace*{0.035truein}
\centerline{\bf BY DIRAC MONOPOLES}
\vspace*{0.035truein}
\vspace*{0.035truein}
\vspace*{0.37truein}
\centerline{\footnotesize YU. P. GONCHAROV}
\vspace*{0.015truein}
\centerline{\footnotesize\it Theoretical Group,
Experimental Physics Department, State Technical University}
\baselineskip=10pt
\centerline{\footnotesize\it Sankt-Petersburg 195251, Russia}
\vspace*{10pt}
\centerline{\footnotesize N. E. FIRSOVA}
\vspace*{0.015truein}
\centerline{\footnotesize\it Institute for Mechanical Engineering,
Russian Academy of Sciences}
\baselineskip=10pt
\centerline{\footnotesize\it Sankt-Petersburg 199178, Russia}
\vspace*{0.225truein}
\pub{13 November 2001}
\vspace*{0.21truein}
\abstracts{
 An algorithm for numerical computation of the barrier transparency for
the potentials surrounding Schwarzschild black holes is described for
massless spinor particles. It is then applied to calculate the all
configurations (including the contributions of twisted field
configurations connected with Dirac monopoles)
luminosity for the Hawking radiation from a Schwarzschild black hole.
It is found that the contribution due to monopoles
can be of order 22~\%  of the all configurations luminosity.
}{}{}
\vspace*{1pt}\textlineskip 
\section{Introductory Remarks} 
\vspace*{-0.5pt}
\noindent
While studying quantum geometry of fields on
black holes the nontrivial topological properties of the latters may play
essential role.
The black holes can actually carry the whole spectrum of topologically
inequivalent configurations (TICs)
for miscellaneous fields, in the first turn, complex scalar and spinor ones.
The mentioned TICs can markedly modify the Hawking radiation from black
holes. Physically, the existence of TICs should
be obliged to the natural presence of magnetic U(N)-monopoles (with $N\ge1$)
on black holes though the total (internal) magnetic charge (abelian or
nonabelian) of black hole remains equal to zero. Up to now, however,
only influence of the TICs of complex scalar field on Hawking radiation
has been studied more or less (see Ref. 1 and
references therein for more details). The description of TICs for
spinors was obtained in Ref. 2 but the detailed analysis of the
TICs contribution to Hawking radiation requires knowledge of the conforming
$S$-matrices which regulate the spinor particle passing through the potential
barrier surrounding black hole. Those $S$-matrices for the Schwarzschild (SW)
black holes have only recently been
explored in Ref. 3 which allow us to obtain an algorithm to
calculate the $S$-matrix elements numerically since for physical results to
be obtained one needs to apply the numerical methods. In the massless case the
$S$-matrices discussed are simpler to treat and the present paper will contain
a description of the algorithm and will apply it to calculate the all
configurations luminosity for massless spinor particles for a SW black hole.
The results obtained can serve as an estimate, in the first turn, of the
electron-positron Hawking radiation and also for the neutrino one from the
SW black holes. The more exact computation should take into account the
particle masses and will require a more complicated algorithm for calculating
the corresponding $S$-matrices. The case of complex scalar field
shows\cite{GF}, however, that the massless limit can be a good approximation
to the more realistic massive case, in particular, when the particles masses
are small enough. As a result, the task under consideration here makes sense.

We write down the black hole
metric under discussion (using the ordinary set of local coordinates
$t,r,\vartheta,\varphi$) in the form
$$ds^2=adt^2-a^{-1}dr^2-r^2(d\vartheta^2+\sin^2\vartheta d\varphi^2) \eqno(1)$$
with $a=1-2M/r$ and $M$ is the black hole mass.

  Throughout the paper we employ the system of units with $\hbar=c=G=1$,
unless explicitly stated otherwise.
Finally, we shall denote $L_2(F)$ the set of the modulo square integrable
complex functions on any manifold $F$ furnished with an integration measure
while $L^n_2(F)$ will be the $n$-fold direct product of $L_2(F)$
endowed with the obvious scalar product.

\section{Description of algorithm}

As was disscussed in Ref. 2,
TICs of a spinor field on black holes are conditioned by the
availability of a countable
number of the twisted spinor bundles over the \bh-topology underlying
the 4D black hole physics. From a physical point of view
the appearance of spinor twisted configurations is linked with the natural
presence of Dirac monopoles that play the role of connections in the
complex line bundles corresponding to the twisted spinor bundles.
Under the circumstances each TIC corresponds to sections of the corresponding
spinor bundle $E$, which can be characterized by its Chern number
$n\in \Bbb{Z}$
(the set of integers).
Using the fact that all the mentioned bundles can be trivilized over
the chart of local coordinates
$(t,r,\vartheta,\varphi) $
covering almost the whole manifold \bh
one can obtain a suitable Dirac equation on the given chart for TIC
$\Psi$
with mass $\mu_0$ and Chern number $n\in\Bbb{Z}$ that looks
as follows
$${\cal D}_n\Psi=\mu_0\Psi,\>\eqno(2)$$
with the twisted Dirac operator ${\cal D}_n=i\gamma^\mu\nabla_\mu^n$
and we can call (standard) spinors corresponding to $n=0$
{\it untwisted} while the rest of the spinors with $n\ne0$
should be referred to as {\it twisted}. Referring for details and for
explicit form of ${\cal D}_n$ to Ref. 2, it should be noted
here that in $L_2^4$(\bh) there is a basis from the solutions of (2)
in the form
$$\Psi_{\lambda m}=\frac{1}{\sqrt{2\pi\omega}}
e^{i\omega t}r^{-1}\pmatrix{F_1(r,\omega,\lambda)
\Phi_{\lambda m}\cr
F_2(r,\omega,\lambda)\sigma_1\Phi_{\lambda m}\cr}\>, \eqno(3)$$
where $\sigma_1$ is the Pauli matrix, the 2D spinor
$\Phi_{\lambda m}=\Phi_{\lambda m}(\vartheta,\varphi)=
(\Phi_{1\lambda m},\Phi_{2\lambda m})$ is
the eigenspinor
of the twisted euclidean Dirac operator with Chern number $n$ on the unit
sphere with the eigenvalue $\lambda =\pm\sqrt{(l+1)^2-n^2}$ while
$-l\le m\le l+1$, $l\ge|n|$. As said above, in the given paper we consider
$\mu_0=0$ and then the functions $F_{1,2}$ obey the system of
equations
$$\cases{\sqrt{a}\partial_rF_1+
\left(\frac{1}{2}\frac{d\sqrt{a}}{dr}+\frac{\lambda}{r}\right)F_1=
-icF_2,\cr
\sqrt{a}\partial_rF_2+
\left(\frac{1}{2}\frac{d\sqrt{a}}{dr}-\frac{\lambda}{r}\right)F_2=
-icF_1\cr} \>\eqno(4)$$
with $c=\omega/\sqrt{a}$ and $a$ of (1). The explicit form of the 2D spinor
$\Phi_{\lambda m}$ is inessential in the given paper and can be found in
Ref. 2. One can only notice here that
they can be subject to the normalization condition at $n$ fixed
$$\int\limits_0^\pi\,\int\limits_0^{2\pi}(|\Phi_{1\lambda m}|^2+
|\Phi_{2\lambda m}|^2)
\sin\vartheta d\vartheta d\varphi=1$$
and these spinors form an orthonormal basis in $L_2^2({\Bbb S}^2)$ at any
$n\in{\Bbb Z}$.

By passing on to the Regge-Wheeler variable $r_*=r+2M\ln(r/2M-1$) and by
going to the quantities $x=r_*/M, y=r/M, k=\omega M$,
we shall have
$x=y+2\ln (0.5y-1)$, so that $y(x)$ is given implicitly by the
latter relation (i.e., $-\infty<x<\infty$, $2\leq y<\infty$) with
$$y'=dy/dx=1-2/y=(y-2)/y=a_0(x)\>\eqno(5)$$
and the system (4) can be rewriten as follows
$$\cases{E'_1 +a_1 E_1=b_1 E_2\>,\cr
 E'_2 + a_2 E_2=b_2E_1\cr} \>\eqno(6)$$
with $E_{1}=E_{1}(x,k,\lambda)=F_+(Mx),
F_+(r^*)=F_{1}[r(r^*)]$,
$E_{2}=E_{2}(x,k,\lambda)=iF_-(Mx),
F_-(r^*)=F_{2}[r(r^*)]$
and
$$a_{1,2}=\frac{1}{2y^2}\pm\frac{\lambda}{y}\sqrt{a_0}\>,\eqno(7) $$
$$b_{1,2}=\mp k \>.\eqno(8) $$
To evaluate luminosity of the Hawking radiation for spinor particles it is
necessary to know the asymptotics of the functions $E_{1,2}$ at $x\to+\infty$.
As was shown in Ref. 3, the latter asymptotics look as follows
$$E_1\sim\sqrt{-k}s_{11}(k,\lambda)e^{ikx},\qquad x\to+\infty \>,\eqno(9)$$
$$E_2\sim\frac{iks_{11}(k,\lambda)}{\sqrt{-k}}e^{ikx},
\qquad x\to+\infty \>,\eqno(10)$$
where $s_{11}(k,\lambda)$ is an element of the $S$-matrix connected with
some scattering problem for the Schr\"odinger-like equation
$$u''+k^2u=qu \>\eqno(11)$$
with potential
$$q(x,\lambda)=\frac{\lambda^2}{y^2(x)}\sqrt{a_0}\left[\sqrt{a_0}+
\frac{y(x)-3}{\lambda y(x)}\right] \>.\eqno(12)$$
In its turn, the correct statement of the mentioned
scattering problem for Eq. (11) consists in searching for two
solutions $u^+(x,k,\lambda)$,
$u^-(x,k,\lambda)$ of the equation (11) obeying the following conditions
$$u^+(x,k,\lambda)=
\cases{e^{ikx}+
s_{12}(k,\lambda)e^{-ikx}+o(1),&$x\to-\infty$,\cr
s_{11}(k,\lambda)e^{ikx}+o(1),&$x\to+\infty$,\cr}$$
$$u^-(x,k,\lambda)=\cases{s_{22}(k,\lambda)e^{-ikx}
+o(1),&$x\to-\infty$,\cr
e^{-ikx}+s_{21}(k,\lambda)e^{ikx}+o(1),&$x\to+\infty$.\cr}\eqno(13)$$
It is seen that there arises some $S$-matrix with elements
$s_{ij}, i, j = 1, 2$. After this one can obtain the luminosity $L(n)$
with respect to the Hawking radiation for TIC
with the Chern number $n$ in the form
(for more details see Ref. 2 and 3)
$$L(n)=
A\sum\limits_{\pm\lambda}\sum\limits_{l=|n|}^\infty2(l+1)
\int\limits_{0}^\infty\,\frac{|s_{11}(k,\lambda)|^2}
{e^{8\pi k}+1}kdk=$$
$$A\sum\limits_{l=|n|}^\infty2(l+1)
\int\limits_{0}^\infty\,\frac{|s_{11}(k,\lambda)|^2+|s_{11}(k,-\lambda)|^2}
{e^{8\pi k}+1}kdk\>,\eqno(14)$$
where
$A=\frac{c^5}{\pi GM}\left(\frac{c\hbar}{G}\right)^{1/2}
\approx0.251455\cdot10^{55}\,
{\rm{erg\cdot s^{-1}}}\cdot M^{-1}$
and $M$ in g.
Luminosity $L(n)$ can be interpreted, as usual,\cite{{GF},{Gon99}}
as an additional contribution to the Hawking radiation due to the additional
spinor particles leaving the black hole because of the interaction with
monopoles and the conforming radiation
can be called {\it the monopole Hawking radiation}.\cite{Gon99}

Under this situation,
for the all configurations luminosity $L$ of black hole with respect
to the Hawking radiation concerning the spinor field to be obtained,
one should sum up over all $n$, i. e.
$$L=\sum\limits_{n\in{\Bbb{Z}}}\,L(n)=L(0)+
2\sum\limits_{n=1}^\infty\,L(n)
\eqno(15)$$
since $L(-n)= L(n)$.

It should be emphasized that in (14), generally speaking,
$s_{11}(k,\lambda)\ne s_{11}(k,-\lambda)$ and that is clear from the form
of potential $q$ of (12) which satisfies the condition\cite{Fir01}
$$\int\limits_{-\infty}^{+\infty}|q(x,\lambda)|dx<\infty\>.$$
As an example, in Figs. 1, 2 the numerically computed potentials
$q(x,\lambda)$ with $\lambda>0$ and $\lambda<0$ are shown.
\begin{figure}[htbp]
\vspace*{13pt}
\centerline{\vbox{\hrule width 5cm height0.001pt}}
\vspace*{1.7truein}
\centerline{\vbox{\hrule width 5cm height0.001pt}}
\vspace*{13pt}
\fcaption{Typical potential barrier for $\lambda>0$.}
\end{figure}

\begin{figure}[htbp]
\vspace*{13pt}
\centerline{\vbox{\hrule width 5cm height0.001pt}}
\vspace*{1.7truein}
\centerline{\vbox{\hrule width 5cm height0.001pt}}
\vspace*{13pt}
\fcaption{Typical potential barrier for $\lambda<0$.}
\end{figure}
One can notice that when $\lambda <0$, $q(x,\lambda)>0$ at any
$x\in[-\infty,\infty]$ while at $\lambda>0$, $q(x,\lambda)$ can change
the sign.

Obviously, for neutrino there exists
only $L(0)$ since neutrino does not interact with monopoles (when neglecting
a possible insignificantly small magnetic moment of neutrino).
Also it is evident that
for numerical computation of all the above luminosities one needs to have some
algorithm for calculating $s_{11}(k,\lambda)$. This can be extracted from
the results of Ref. 3. Namely
$$s_{11}(k,\lambda)=2ik/[f^-(x,k,\lambda),f^+(x,k,\lambda)]\>,\eqno(16)$$
where [,] signifies the Wronskian of functions $f^-,f^+$, the
so-called Jost type solutions of Eq. (11).
In their turn, these functions and
their derivatives obey the certain integral equations. Since the Wronskian
does not depend on $x$ one can take the following form of the mentioned
integral equations

$$f^-(x_0,k,\lambda)=e^{-ikx_0}+\frac{1}{k}\int\limits^{x_0}_{-\infty}
\sin[k(x_0-t)]q(t,\lambda)f^-(t,k,\lambda)dt\>,\eqno(17)$$
$$(f^-)'_x(x_0,k,\lambda)=-ike^{-ikx_0}+
\int\limits^{x_0}_{-\infty}\cos[k(x_0-t)]q(t,\lambda)f^-(t,k,\lambda)dt\>,
\eqno(18)$$
$$f^+(x_0,k,\lambda)=e^{ikx_0}+\frac{1}{k}
\int\limits_{x_0}^{+\infty}\sin[k(x_0-t)]
q(t,\lambda)f^+(t,k,\lambda)dt\>,\eqno(19)$$
$$(f^+)'_x(x_0,k,\lambda)=ike^{ikx_0}+\int\limits^{+\infty}_{x_0}
\cos[k(x_0-t)]q(t,\lambda)f^+(t,k,\lambda)dt\>.\eqno(20)$$
The point $x_0$ should be chosen from the considerations of
the computational convenience.
The relations (17)--(20) can be employed for numerical calculation of
$s_{11}(k,\lambda)$.

Finally, we should touch upon the convergence of the
series (15) over $n$.
For this aim we denote
$$c_l(\pm\lambda)=
\int\limits_{0}^\infty\,\frac{|s_{11}(k,\pm\lambda)|^2}
{e^{8\pi k}+1}kdk$$
and we represent the coefficients $c_l(\pm\lambda)$ in the form
(omitting the integrand)
$$c_l(\pm\lambda)=c_{l1}(\pm\lambda)+c_{l2}(\pm\lambda)=
\int\limits_0^{(\ln l)/2\pi}+\int\limits_{(\ln l)/2\pi}^\infty
\>,\eqno(21)$$
so that $L(n)$ of (14) is equal to $L_1(n)+L_2(n)$ respectively. Also it
should be noted, as follows from the results of Ref. 3, we
have for barrier transparency $\Gamma(k,\lambda)=|s_{11}(k,\lambda|^2$  that
$\Gamma(0,\lambda)=0$ and at $k\to+\infty$
$$|s_{11}(k,\lambda)|^2=\Gamma(k,\lambda)=1+O(k^{-1})\>.\eqno(22)$$
Further we have the asymptotic behaviour for large $l$ and $k<<l$
$$|s_{11}(k,l,n)|= C\frac{e^{-\sqrt{n+l+1}}}{\sqrt{n+l+1}}k[1+o(1)]\>
\eqno (23)$$
with some constant $C$. To obtain it we should work using the equations
(17)--(20) by conventional methods
of mathematical analysis so that the detailed derivation of (23) lies somewhat
out of the scope of the present paper and will be considered elesewhere.
Using (21) and (23) we find
$$c_{l1}(\pm\lambda)\leq B\frac{e^{-2\sqrt{n+l+1}}}{n+l+1}
\int_0^{(\ln l)/2\pi}\frac{k^3}{e^{8\pi k}+1}dk$$
$$\leq B\frac{e^{-2\sqrt{n+l+1}}}{n+l+1}
\int_0^\infty\frac{k^3}{e^{8\pi k}-1}dk=
B\frac{e^{-2\sqrt{n+l+1}}}{n+l+1}\frac{3! \zeta(4)}{(8\pi)^4}\>$$
with some constant $B$,
where we used the formula (for natural $p>0$)\cite{Abr64}
$$\int\limits_0^\infty{t^pdt\over e^t-1}=p!\,\zeta(p+1)$$
with the Riemann zeta function $\zeta(s)$, while $\zeta(4)=\pi^4/90$.
As a consequence, one may consider
$$ L_1(n)\sim\int\limits_{n}^{\infty}2(l+1)\frac{e^{-2\sqrt{n+l+1}}}{n+l+1}
dl\sim2\int\limits_{\sqrt{2n+1}}^\infty e^{-2t}\left(t-\frac{n}{t}\right)dt$$
$$\sim e^{-2\sqrt{2n+1}}\left(\sqrt{2n+1}-\frac{1}{\sqrt{2n+1}}\right)
\>, \eqno (24)$$
where we employed the asymptotical behaviour of the incomplete gamma
function\cite{Abr64}
$$\Gamma(\alpha,x)=\int\limits_x^\infty e^{-t}t^{\alpha-1}dt
\sim x^{\alpha-1}e^{-x},\>x\to+\infty\>,$$
so that the series $\sum_1^\infty L_1(n)$ is evidently convergent.

At the same time due to (22)
$$c_{l2}(\pm\lambda)\leq D\int\limits_{(\ln l)/2\pi}^\infty ke^{-8\pi k}dk
\sim D\frac{\ln l}{l^4}\eqno(25)$$
with some constant $D$. This entails
$$L_2(n)\sim \int\limits_{n}^\infty\frac{2(l+1)\ln l}{l^4}dl\sim
\frac{\ln n}{n^2}\>.\eqno(26)$$
and the series $\sum_1^\infty L_2(n)$ is also convergent.
Under this situation we obtain that
the series of (15) is convergent, i. e., $L<+\infty$.
Consequently, we can say that all the further computed
luminosities are well defined and exist.

\section{Numerical results}
 In view of (22), we have
$$\int\limits_{0}^{\infty}\frac{\Gamma(k,\lambda)kdk}
{e^{8\pi k}+1}\sim
\int\limits_0^\infty\,\frac{kdk}{e^{8\pi k}-1}=\frac{1}{(8\pi)^2}\zeta(2)=
\frac{1}{6\cdot8^2}\>,\eqno(27)$$
whilst the latter integral can be accurately evaluated using the
trapezium formula on the [0, 3] interval. Having confined the range of $k$
to the [0, 3] interval, accordingly, it is actually enough
to restrict oneself to $0\leq l,n\leq$15--20 when computing $L(n),L$.
We took $0\leq l\leq20$, $0\leq n\leq15$.
Since the Wronskian of (16) does not depend on $x$, the latter should be
chosen in the region where the potentials $q(x,\pm\lambda)$ are already
small enough. Besides,
potentials $q$ are really equal to 0 when $x<$ -30. We computed
$\Gamma(k,\lambda)$ according to (16) at $x=x_0$ = 300,
where $f^\pm$ were obtained
from the Volterra integral equations, respectively, (17) and (19). For this
aim, according to the methods developed for numerical solutions of integral
equations (see e. g., Ref. 5),
those of (17) and (19) have been replaced by systems of linear algebraic
equations which can be gained when calculating the conforming integrals by
the trapezium formula, respectively, for the [-30, 300] and [300, 400]
intervals. The sought values of $f^\pm$ were obtained as the solutions to the
above linear systems.
After this the derivatives $(f^\pm)'_x$ were evaluated in accordance with
(18) and (20) while employing the values obtained for $f^\pm$. The typical
behaviour of $\Gamma(k,\lambda)$ is presented in Fig. 3.

\begin{figure}[htbp]
\vspace*{13pt}
\centerline{\vbox{\hrule width 5cm height0.001pt}}
\vspace*{1.7truein}
\centerline{\vbox{\hrule width 5cm height0.001pt}}
\vspace*{13pt}
\fcaption{Typical behaviour of barrier transparencies.}
\end{figure}

Finally, Fig. 4 represents the untwisted $L(0)/A$ and the all
configurations
$L/A$ luminosities with $A$ of (14) as functions
of $k$. The areas under the curves give the corresponding values
of $L(0)/A$ and $L/A$
  $$L(0)/A = 0.514490\cdot10^{-3}\>,$$
$$   L/A= 0.658571\cdot10^{-3}\>,$$
$$  L(0)/L= 0.781222 \>,$$
so that contribution owing to Dirac monopoles amounts to
$21.8778 \% \sim 22 \%$ of $L$. For inquiry, the similar contribution
for scalar particles was of order (17--18) \%.\cite{GF}

\begin{figure}[htbp]
\vspace*{13pt}
\centerline{\vbox{\hrule width 5cm height0.001pt}}
\vspace*{1.7truein}
\centerline{\vbox{\hrule width 5cm height0.001pt}}
\vspace*{13pt}
\fcaption{Untwisted and all configurations luminosities.}
\end{figure}

\section{Concluding remarks}
As was mentioned above, the results obtained can serve as an estimate,
in the first turn, of the electron-positron Hawking radiation and also
of that for neutrino from the SW black holes.
Clearly, for neutrino there exists
only $L(0)$ since neutrino does not interact with monopoles (when neglecting
a possible insignificantly small magnetic moment of neutrino).
The corresponding computation for other fundamental fermions
($\mu^{\pm}$-mesons and $\tau^{\pm}$-leptons)
and also the more exact computation for $e^{\pm}$-particles should take into
account the particle masses and will require a more complicated algorithm
for calculating the corresponding $S$-matrices.
We hope to get the numerical results in the massive case elsewhere.

\nonumsection{Acknowledgements}
\noindent
    The work of Goncharov was supported in part by the Russian Foundation for
Basic Research (grant No. 01--02--17157).
\newpage
\nonumsection{References}

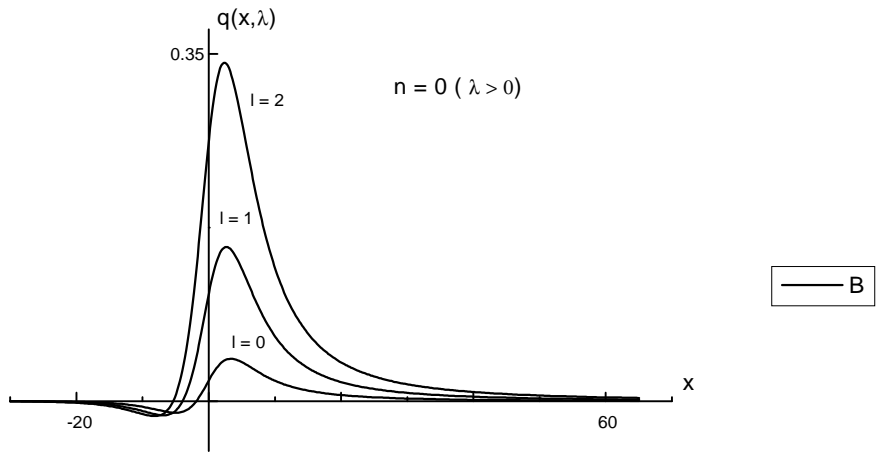
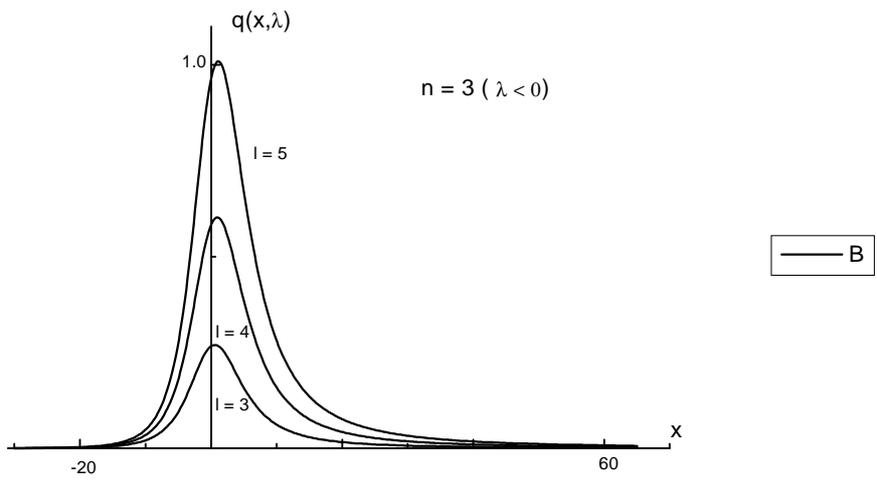
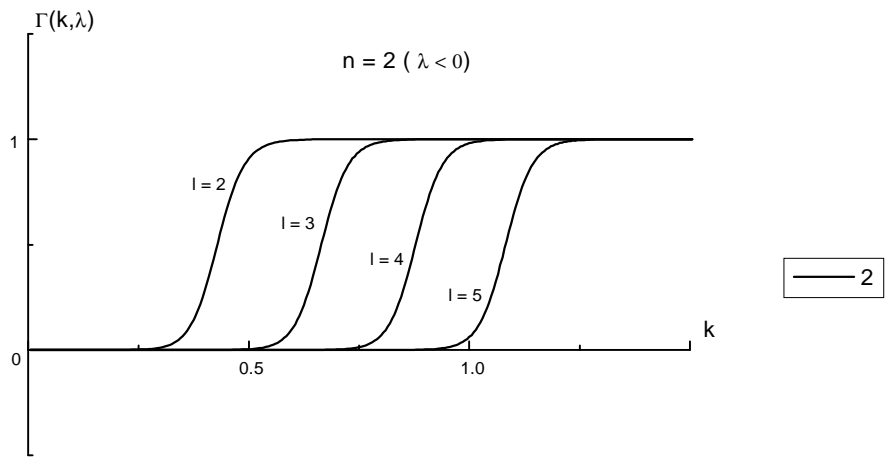
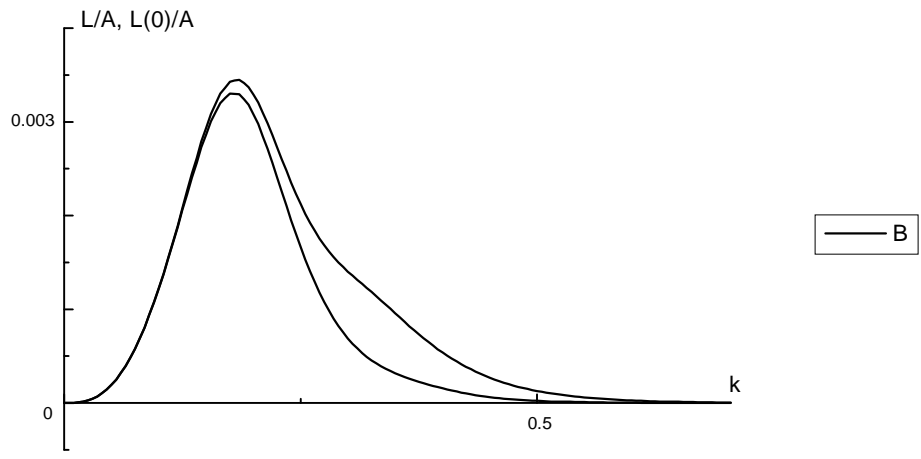
\end{document}